\documentclass[seceq]{ptptex}
\usepackage{amsmath,amssymb}
\usepackage[dvips]{graphicx}

\title{
A Method to Construct Asymptotic Solutions \\
Invariant under the Renormalization Group

}

\author{
Masatomo \textsc{IWASA}\footnote{e-mail:miwasa@.phys.nagoya-u.ac.jp} and Kazuhiro \textsc{NOZAKI}\footnote{e-mail:knozaki@r.phys.nagoya-u.ac.jp}%
}

\inst{
Departmant of Physics, Nagoya University, Nagoya 464-8602, Japan
}

\abst{
     A renormalization group method with the Lie symmetry is presented  
for
the singular perturbation
     problems. Asymptotic solutions  are obtained  as group-invariant
solutions under approximate Lie groups admitted by perturbed diferential
equations.
}

\begin{document}

\maketitle

\section{Introduction}

There have been many studies concerning application of the renormalization group theory
of particle physics as a singular perturbation method to treat 
differential equations since the work of the Illinois group\cite{IllinoisRG}.
In this method, integral constants appearing in the lowest-order
perturbed solution are renormalized in order to
remove secular or divergent terms appearing in the naive perturbation
solution and give a well-behaved asymptotic solution. We call this
renormalization group method in 
singular perturbation theory the ``conventional RG method'' in this paper.
A geometrical aspect of the conventional RG method has been studied by means
of the envelopes of solutions  \cite{Kunihiro}. The conventional RG
method and
its variants depend more or less on the naive perturbation analysis.
In connection to this, it is interesting that the geometrical symmetry  of a differential equation
with respect to
continuous transformations, i.e. the Lie symmetry, is known to be
useful for deriving new solutions from
known solutions \cite{Olver}. There have been some pioneering works
aimed of  
constructing the renormalization
group method in terms of the Lie symmetry. \cite{Shirkov1}
\cite{Shirkov2} However, to this time, there is no case in which
have been derived
asymptotic solutions in singular perturbation problems by means of
the Lie symmetry.  The purpose of this paper is to derive such 
asymptotic solutions in the framework of the Lie group and symmetry.

\section{Method of the renormalization group with symmetry}
        \label{sec2}
        For simplicity, let us consider the second-order ordinary
        differential equation
        \begin{eqnarray}
         F_0(u,u',u'')+\varepsilon F_1(u,u') = 0,
         \label{eq:2-1}
        \end{eqnarray}
        where we have $u'=\frac{du}{dt}$ and $u''=\frac{d^2 u}{dt^2}$,
	$F_0$ and $F_1$ are
        some functions of their arguments, and $\varepsilon$ is a small
	parameter.  
        This system is assumed to be nearly solvable in the sense that
	the general solution is known for the unperturbed case,
	$\varepsilon = 0$.   
        Our purpose is to construct an approximate solution of the system
        (\ref{eq:2-1}) for small $\varepsilon$ by incorporating an
        approximate Lie symmetry.
        First, the equation (\ref{eq:2-1}) is rewritten as a system of
	first-order ordinary differential equations in the form
        \begin{eqnarray}
         \left\{
              \begin{array}{l}
                u'=v ,\\
                F_0(u,v,v') +\varepsilon F_1(u,v)=0 ,
                \label{eq:2-2}
              \end{array}
         \right.
        \end{eqnarray}
       or as a first-order ordinary differential equation with
       respect to a complex valuable $z:=u+iv$ as
        \begin{eqnarray}
         G_0(z,\bar{z},z',{\bar{z}}')+\varepsilon G_1(z,\bar{z}) = 0.
          \label{eq:2-2.1}
        \end{eqnarray}
        Let (\ref{eq:2-2}) admits a Lie group transformation whose
        infinitesimal generator takes the form
        \begin{eqnarray}
           X = \partial_\varepsilon
                      +\zeta(t,u,v)\partial_t
                      +\eta^u(t,u,v)\partial_u
                      +\eta^v(t,u,v)\partial_v,
            \label{eq:2-2.5}
        \end{eqnarray}
        Then, under the transformation $(u,v)\mapsto(z,\bar{z})$,
        (\ref{eq:2-2.5}) can be rewritten as
        \begin{eqnarray}
           X = \partial_\varepsilon
                +\xi(t,z,\bar{z})\partial_t
                +\eta^z(t,z,\bar{z})\partial_z
                +\eta^{\bar{z}}(t,z,\bar{z})\partial_{\bar{z}},
             \label{eq:2-3}
        \end{eqnarray}
        where
        \begin{eqnarray}
          \xi = \bar{\xi} = \zeta, \ \
           \eta^z = \eta^u+i\eta^v,\ \
           \eta^{\bar{z}} = \eta^u-i\eta^v
                          = \overline{\eta^z}. \ \
        \end{eqnarray}
        Next, let
        \begin{eqnarray}
         X^{(1)}(t,z,\bar{z},z',{\bar{z}}')
           &=& \partial_\varepsilon
              +\xi(t,z,\bar{z})\partial_t
              +\eta^z(t,z,\bar{z})\partial_z
              +\eta^{\bar{z}}(t,z,\bar{z})\partial_{\bar{z}}
            \nonumber\\
           && +\eta^{z(1)}(t,z,\bar{z},z',{\bar{z}}')\partial_{z'}
+\eta^{\bar{z}(1)}(t,z,\bar{z},z',{\bar{z}}')\partial_{\bar{z}'}
            \label{eq:2-4}
        \end{eqnarray}
        be the prolongation of X.
           Then, the determining equation for (\ref{eq:2-2.1}), which
           determines each component of the vector field (\ref{eq:2-3}),
           is given by
           \begin{eqnarray}
	       X^{(1)}\{G_0(z,\bar{z},z',{\bar{z}}')+\varepsilon
G_1(z,\bar{z})\}\Bigm|_
		{
		{\scriptsize
			       G_0+\varepsilon G_1=0
                     }
		}              =0.
                  \label{eq:2-5}
	    \end{eqnarray}
      	  Because we wish to find the approximate symmetries to leading
	   order, we need only  
	   solve the following leading-order determining  equation:
	  \begin{eqnarray}
	     X^{(1)}\{G_0(z,\bar{z},z',{\bar{z}}')
                  +\varepsilon G_1(z,\bar{z})\}\Bigm|_
	         {
	         {\scriptsize 
			     G_0=0
                    }
	      }              =O(\varepsilon) .
                \label{eq:2-6}
	  \end{eqnarray}
	  By solving the approximate determining equation, (\ref{eq:2-6}),  we
obtain the infinitesimal generator $X$   admitted by
the system (\ref{eq:2-2.1})
in the leading-order approximation.	 As shown below in some examples,
the approximate solutions of the determining equation (\ref{eq:2-6})
give various non-trivial
approximate symmetries even if the system (\ref{eq:2-2.1}) admits only a
few exact, trivial symmetries. It is by virtue of this fact that the present
method is useful in obtaining approximate solutions. Using the
approximate infinitesimal
generator $X$, we construct  a group-invaritant solution of the system
(\ref{eq:2-2.1}).

The group-invaritant solution,
               $z=z(\varepsilon,t)$ , which
satisfies
	  \begin{eqnarray}
	       X\{z-z(\varepsilon,t)\}\Bigm|_
		{
		{\scriptsize 
			       z-z(\varepsilon,t)=0
                  }
                  }
                     =0,
               \end{eqnarray}
is obtained by solving the so-called Lie equation
               \begin{eqnarray}
	      \partial_\varepsilon z
                     =-\xi(t,z,\bar{z}) \partial_t z
                     +\eta^z(t,z,\bar{z}).
                \label{eq:2-7}
	  \end{eqnarray}
Because our symmetry (\ref{eq:2-3}) always has a non-vanishing
component in the  $\varepsilon$ direction, the  Lie equation becomes a
system of differential equations with respect to the perturbation
parameter $\varepsilon$ as described above. Noting this fact, we refer
to the Lie
equation (\ref{eq:2-7}) as the renormalization group
equation.
      Solving the renormalization group equation by adopting solutions of
	the unperturbed system as boundary condition, i.e.
               \begin{eqnarray}
                  z(\varepsilon=0,t)=z^{(0)}(t)\ \ \
	   s.t.\ \ \
G_0(t,z^{(0)},{\overline{z^{(0)}}},{z^{(0)}}',{\overline{z^{(0)}}}')=0,
               \label{eq:2-8}
              \end{eqnarray}
                we obtain an approximate solution of the system
(\ref{eq:2-2.1})
for non-vanishing $\varepsilon$,
	  \begin{eqnarray}
	     z=z(\varepsilon,t).
                \label{eq:2-9}
	  \end{eqnarray}
    As the solution thus obtained  is invariant  with respect to
the Lie group,
    we call it the invariant renormalized solution.

       Here, some detailed discussion is necessary to regorously construct our
        remormalization group method with the Lie symmetry.  In the
        above-described general procedure, 
        we obtain the approximate symmetry only to leading order in
        $\varepsilon$, that is, $\varepsilon^0$. Nevertheless, the
        solution invariant with respect to the leading-order
	Lie symmetry yields an approximate solution up to
        $O(\varepsilon)$ of the system (\ref{eq:2-2.1}).
    Is this actually a solution of the given differential equations
    to order $\varepsilon^1$?  The following proposition answers
this question.\\
{\bf Proposition. }
        Consider a system of ordinary differential equations of the form
        \begin{eqnarray}
            G_0(z,\bar{z},z',{\bar{z}}')+\varepsilon G_1(z,\bar{z})=0.
         \label{eq:2-10}
        \end{eqnarray}
        Assume that (\ref{eq:2-10}) approximately admits a Lie group
        transformation with an infinitesimal generator  $X$, i.e. 
        \begin{eqnarray}
            X^{(1)}\{G_0(z,\bar{z},z',{\bar{z}}')
                     +\varepsilon G_1(z,\bar{z})\}\Bigm|_{
                {\scriptsize
                      G_0=0
                }
                }
              =O(\varepsilon),
         \label{eq:2-11}
        \end{eqnarray}
        where $X^{(1)}$ is the first prolonged infinitesimal generator.
        Then, its invariant solution is a
        solution of (\ref{eq:2-10}) to order $\varepsilon^1$\\
        {\bf Proof.}
         Suppose that (\ref{eq:2-10}) admits
         \begin{eqnarray}
          X=\partial_\varepsilon
                         +\xi(t,z,\bar{z})\partial_t
                         +\eta^z(t,z,\bar{z})\partial_z
                         +\eta^{\bar{z}}(t,z,\bar{z})\partial_{\bar{z}}.
          \label{eq:2-12}
         \end{eqnarray}
         Then its invariant solution is a solution of
         \begin{eqnarray}
             && X\{z-z(\varepsilon,t)\}\Bigm|_{
                {\scriptsize
                    z=z(\varepsilon,t)
                }
               }=0\\
          &&\Longleftrightarrow
	 \partial_\varepsilon z(\varepsilon,t)
             =-\xi(t,z,\bar{z})\partial_t
z(\varepsilon,t)+\eta^z(t,z,\bar{z}),
v(\varepsilon,t)=-\xi(t,u,v)\partial_t
          \label{eq:2-13}
         \end{eqnarray}
        Thus, the invariant solution can be expressed as
         \begin{eqnarray}
           z(\varepsilon,t)
             &=&z^{(0)}(t)
                +\int_{0}^{\varepsilon}\left\{-\xi(t,z,\bar{z})
                \partial_t z(\varepsilon,t)
                +\eta^z(t,z,\bar{z})\right\}d\varepsilon 
                     \label{eq:2-13.9}\\
             &=&z^{(0)}
-\int_{0}^{\varepsilon}\left(\xi(t,z^{(0)},{\bar{z}}^{(0)})
                     z^{(0)'}
                -\eta^{z}(t,z^{(0)},{\bar{z}}^{(0)})\right)d\varepsilon
              +O(\varepsilon^2)\\
             &=&z^{(0)}-\varepsilon\left(\xi(t,z^{(0)},{\bar{z}}^{(0)})z^{(0)'}
                -\eta^z(t,z^{(0)},{\bar{z}}^{(0)})\right)
              +O(\varepsilon^2),
                             \label{eq:2-14}
          \end{eqnarray}
         where $z^{(0)}(t)$ and $\bar{z}^{(0)}(t)$ are the solutions of
the
         unperturbed system. By substituting (\ref{eq:2-14})
	into (\ref{eq:2-10}) and taking account of (\ref{eq:2-11}), we
	can show that the solutions satisfy the given 
	differential equations to order $\varepsilon^1$  $\Box$

     It is noted that this proposition is easily proved for a solution
     of arbitarary order in $\varepsilon$ by expanding (\ref{eq:2-13.9})
     to that order.  
     Owing to this proposition, it is not necessary to know a naive
        perturbed solution in order to find the proper Lie symmetry.
     This contrasts with the situation in the RG method of Shirkov and
     Kovalev, \cite{Shirkov1} \cite{Shirkov2} in which the naive
     perturbed solution is needed. 

\section{Examples}
        \label{sec3}

Let us consider two examples. \\
       (1) {\it a Harmonic oscillator equation        }\\
        \label{sec2.2}
        Here, we apply the renormalization group method
        with the Lie symmetry to the equation for a
        harmonic oscillator as an example of
        singular perturbaiton problems. We consider the equation
        \begin{eqnarray}
         u''+u=-\varepsilon u.
         \label{eq:2-16}
        \end{eqnarray}
        Then, defining $z:=u+iu'$, we have
        \begin{eqnarray}
           z'=-iz-\varepsilon \frac{1}{2}i\left( z+\bar{z} \right)
           \label{eq:2-17}
        \end{eqnarray}
        Next, let 
        \begin{eqnarray}
           X(t,z,\bar{z})=\partial_{\varepsilon}+\xi(t,z,\bar{z})\partial_t
+\eta^z(t,z,\bar{z})\partial_z+\eta^{\bar{z}}(t,z,\bar{z})\partial_{\bar
{z}}
         \label{eq:2-18}
        \end{eqnarray}
        be the infinitesimal generator of an approximate Lie symmetry
	admitted by (\ref{eq:2-17}).
        Then, we write its first prolonged infinitesimal generator as
        \begin{eqnarray}
      X^{(1)}(t,z,\bar{z},z',\bar{z}')&
           =&\partial_{\varepsilon}
            +\xi(t,z,\bar{z})\partial_t
            +\eta^z(t,z,\bar{z})\partial_z
            +\eta^{\bar{z}}(t,z,\bar{z})\partial_{\bar{z}}\nonumber\\
&&\hspace{1.0cm}
            +{\eta^{z}}^{(1)}(t,z,\bar{z},z',\bar{z}')\partial_{z'}
            +{\eta^{\bar{z}}}^{(1)}(t,z,\bar{z},z',\bar{z}')\partial_{\bar{z}'},
         \label{eq:2-19}
        \end{eqnarray}
       where
        \begin{eqnarray}
&&\hspace{-0.5cm}{\eta^z}^{(1)}(t,z,\bar{z},z',\bar{z}')
                    =\eta_t^z
                     +z'(\eta^z_z-\xi_t)
                     +\bar{z}'\eta^z_{\bar{z}}
                     -{z'}^2\xi_z
                     -z'\bar{z}'\xi_{\bar{z}},
           \label{eq:2-20}\\
&&\hspace{-0.5cm}{\eta^{\bar{z}}}^{(1)}(t,z,\bar{z},z',\bar{z}')
                    =\eta_t^{\bar{z}}
                     +\bar{z}'(\eta^{\bar{z}}_{\bar{z}}-\xi_t)
                     +z'\eta^{\bar{z}}_z
                     -\bar{z}'^2 \xi_{\bar{z}}
                     -z'\bar{z}'\xi_z.
           \label{eq:2-21}
        \end{eqnarray}
         Here, the subscript
	$\alpha\ (\alpha=t,u,v)$ of $\xi$and $\eta$ represents the operation of
	$\partial_\alpha$.
       Then, the approximate determining equation for (\ref{eq:2-17}) is
        \begin{eqnarray}
            X^{(1)}\left\{z'+iz+\varepsilon
\frac{1}{2}i\left(z+\bar{z}\right)\right\}
             \Bigm|_{
                 {\scriptsize
                   z'+iz=0
	          }}
            =O(\varepsilon).
          \label{eq:2-22}
         \end{eqnarray}
       This equation leads to
        \begin{eqnarray}
           {\eta^z}^{(1)}\Bigm|_
             {z'+iz=0}
             +i\eta^z
             +\frac{1}{2}i\left(z+\bar{z}\right)=0.
        \label{eq:2-24}
       \end{eqnarray}
     The system (\ref{eq:2-24}) is  linear in the unknown $\eta^z$ and  
$\xi$
     with an inhomogeneous term due to the perturbation. Because we are
     interested in the effect of the perturbation, we seek a particular
     solution of the linear inhomogeneous system (\ref{eq:2-24}).
     Expanding $\xi$ and $\eta^u$ in powers of $t, z$ and $\bar{z}$ as
        \begin{eqnarray}
          \xi(t,z,\bar{z}) = \sum_{j,k,l \geq 0}\xi_{jkl}t^j z^k
\bar{z}^l,
        \label{eq:2-26}\\
          \eta^z(t,z,\bar{z})= \sum_{j,k,l \geq 0}\eta_{jkl}t^j z^k
\bar{z}^l.
        \label{eq:2-27}
        \end{eqnarray}
      and substituting these expressions into (\ref{eq:2-24}),
       we obtain a particular solution of lowest order,
       \begin{eqnarray}
         &\xi(t,z,\bar{z})   &=0,
            \label{eq:2-31-1}\\
         &\eta^z(t,z,\bar{z})&=-\frac{1}{2}itz+\frac{1}{4}z-\frac{1}{4}\bar{z}.
            \label{eq:2-31-2}
        \end{eqnarray}
        Then,  the infinitesimal generator admitted approximately by
(\ref{eq:2-17}) is
        \begin{eqnarray}
          X(t,u,v)=\partial_\varepsilon
                   +\left(-\frac{1}{2}itz+\frac{1}{4}z-\frac{1}{4}\bar{z}\right)\partial_z
                  +\left(\frac{1}{2}it\bar{z}-
\frac{1}{4}z+\frac{1}{4}\bar{z}\right)\partial_{\bar{z}}.
           \label{eq:2-32}
        \end{eqnarray}
      The invariant solution is derived from the corresponding Lie
equation
       or  renormalization group equation, and we have
       \begin{eqnarray}
	    \partial_\varepsilon z&
                  =&-\frac{1}{2}itz+\frac{1}{4}z-\frac{1}{4}\bar{z}.
	    \label{eq:2-40}
	 \end{eqnarray}
             This renormalization group equation is consistent with the
	naive expanded solution of
	(\ref{eq:2-17}),
       \begin{eqnarray}
             z(\varepsilon,t)&
               =&Ae^{-i(t+\theta)}
\hspace{-1mm}
                +
\hspace{-1mm}
                \varepsilon
                \left(
\hspace{-1mm}
                -\frac{1}{2}iAte^{-i(t+\theta)}
                +\frac{1}{4}Ae^{-i(t+\theta)}
                -\frac{1}{4}\bar{A}e^{i(t+\theta)}
\hspace{-1mm}
                \right)
\hspace{-1mm}
                +O(\varepsilon^2).
               \label{eq:2-41}
	\end{eqnarray}
          Then, defining $z_0:=Ae^{-i(t+\theta)}$,
          Eq.(\ref{eq:2-41}) reads
	\begin{eqnarray}
	   z(\varepsilon,t)
                      =z_0
                      +\varepsilon\left(
                      -\frac{1}{2}itz_0
                      +\frac{1}{4}z_0
                      +\frac{1}{4}\bar{z}_0
                      \right),
              \label{eq:2-43}
	\end{eqnarray}
	which is the Euler scheme difference equation corresponding
	to the renormalization equation (\ref{eq:2-40}).
          Using the unperturbed solution as the boundary condition, i.e.
            \begin{eqnarray}
               z(\varepsilon=0,t)=Ae^{-i(t+\theta)},
            \end{eqnarray}	
	we obtain the invariant renormalized solution
	\begin{eqnarray}
	 u(\varepsilon,t)&=&2\left(t^2-\frac{1}{4}\right)^{-\frac{1}{2}}
           e^{\frac{\varepsilon}{4}}R
           \sin\left(\frac{1}{2}\left(t^2-\frac{1}{4}\right)
           ^{\frac{1}{2}}\varepsilon\right)
      	  \left(2t\cos(t+\theta)-\sin(t+\theta)\right)\nonumber \\
	 &&+e^{\frac{\varepsilon}{4}}R
	  \cos\left(\frac{1}{2}\left(t^2-\frac{1}{4}\right)^{\frac{1}{2}}
	       \varepsilon\right)\sin(t+\theta),
            \label{eq:2-44}
	\end{eqnarray}
      where $R = |A|$. 
       Although the expression (\ref{eq:2-44}) is somewhat complicated,
       it is easy to
       show that this invariant solution asymptotically approaches the
       renormalized solution derived using
the conventional RG method. Since we are interested in the long-time
behavior of solution [i.e.
$t\sim (1/\varepsilon)$ in this case] we set $(t^2-\frac{1}{4})\approx
t^2$ and $e^{\frac{\varepsilon}{4}}\approx 1$ and ignore
$\sin(t+\theta)$  in the first term. Then, the
solution (\ref{eq:2-44}) reduces to the conventional renormalized
solution,
       \begin{eqnarray}
         u(\varepsilon,t)
           =R\sin\left(\left(1+\varepsilon\frac{1}{2}\right)t+\theta\right).
       \end{eqnarray}

      Adding the general solution of the determining equation
      (\ref{eq:2-24}) to the particular solution given in Eqs.(\ref{eq:2-31-1}) and
      (\ref{eq:2-31-2}), we can construct a simpler invariant  
renormalized
      solution. For example, choose a particular solution of
      (\ref{eq:2-24}) as
       \begin{eqnarray}
          &\xi(t,u,v)   &=-\frac{1}{2}t,
            \label{eq:2-34-1}\\
          &\eta^z(t,u,v)&=\frac{1}{4}(z-\bar{z}).
            \label{eq:2-34-2}
        \end{eqnarray}
      Then, the Lie equation becomes
      	\begin{eqnarray}
	 \partial_\varepsilon z
              =\frac{1}{2}t\partial_t z
               +\frac{1}{4}(z-\bar{z}) .
              \label{eq:2-35}
	\end{eqnarray}
Thus, we have the simpler invariant renormalized solution
       	\begin{eqnarray}
	 u(\varepsilon,t)=R\sin(e^{\frac{\varepsilon}{2}}t+\theta),
              \label{eq:2-39}
	\end{eqnarray}
which also is identical with the conventional renormalized solution to
order $\varepsilon^1$. 
These two invariant renormalized solutions are depicted
in Fig. 1 along with the exact solution, the naive perturbation solution and
the conventional renormalized solution.
We observe good agreement among these solutions even for $\varepsilon
=0.2$, except for the naive  perturbation solution.

       \begin{figure}
       \begin{center}
        \rotatebox{0}{
\includegraphics[width=8cm,height=12cm,keepaspectratio]{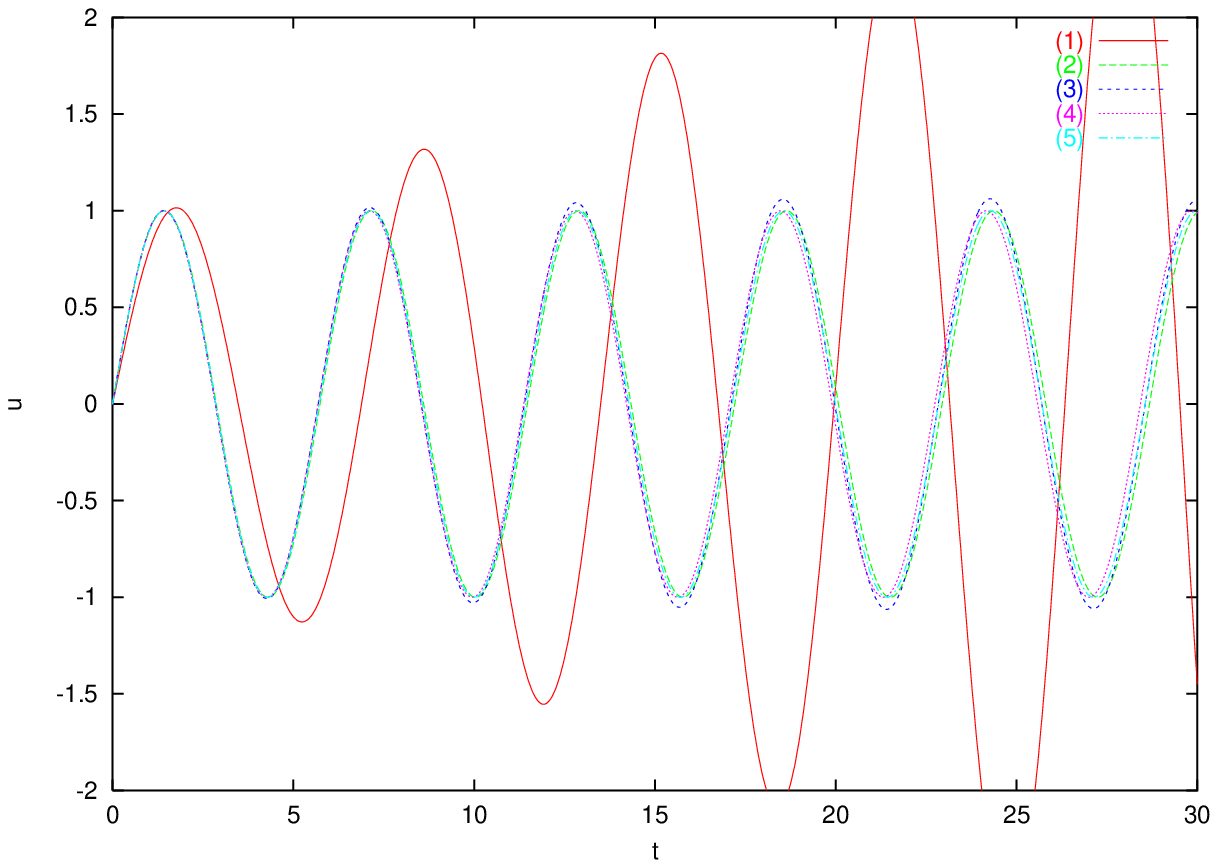}}
       \end{center}
       \caption{Comparison of perturbative solutions of
	Eq.(\ref{eq:2-16}). (1) represents the naive perturbation
	solution of order $\varepsilon$, (2) the exact solution, (3) and
	(4) the invariant renormalizad solutions (\ref{eq:2-44}) and
	(\ref{eq:2-39}), respectively, and (5) conventional renormalized
	solution, where $\theta=0,R=1$,and $\varepsilon=0.2$}
       \label{fig:linear}
      \end{figure}
(2) {\it Rayleigh equation}\\
We next consider the Rayleigh equation,
      \begin{eqnarray}
        u''+u=\varepsilon\left(u'-\frac{1}{3}u'^3\right),
       \label{eq:2-46}
      \end{eqnarray}
which can be rewitten
      \begin{eqnarray}
        && z'=-iz
            +\varepsilon\left(\frac{1}{2}z-\frac{1}{2}\bar{z}
               +\frac{1}{24}z^3-\frac{1}{8}|z|^2z
               +\frac{1}{8}|z|^2\bar{z}-\frac{1}{24}\bar{z}^3\right).
\end{eqnarray}
In this case, the determining equation for the components
$(\xi,\eta^z,\eta^{\bar{z}})$ of the infinitesimal generator is
        \begin{eqnarray}
       {\eta^z}^{(1)}\Bigm|_{z'+iz=0}+i\eta^z
             +\frac{1}{2}z-\frac{1}{2}\bar{z}
             +\frac{1}{24}z^3-\frac{1}{8}|z|^2z
             +\frac{1}{8}|z|^2\bar{z}-\frac{1}{24}\bar{z}^3=0.
        \label{eq:2-47}
       \end{eqnarray}
Following the same procedure as in the previous example,
we obtain the following particular solution of (\ref{eq:2-47}):
       \begin{eqnarray}
         &\xi(t,u,v)   &=0,
            \label{eq:2-48-1}\\
         &\eta^z(t,z,\bar{z})&
            =\frac{1}{4}i\bar{z}+\frac{1}{2}tz
             +\frac{1}{48}iz^3-\frac{1}{96}i\bar{z}^3
             -\frac{1}{16}i|z|^2\bar{z}-\frac{1}{8}t|z|^2z.
            \label{eq:2-48-2}
        \end{eqnarray}
From this, we find the infinitesimal generator approximately admitted by
Eq.(\ref{eq:2-46}) to be
\begin{eqnarray}
         X(t,u,v)&
           =&\partial_\varepsilon
             +\left(\frac{1}{4}i\bar{z}+\frac{1}{2}tz
             +\frac{1}{48}iz^3-\frac{1}{96}i\bar{z}^3
             -\frac{1}{16}i|z|^2\bar{z}
             -\frac{1}{8}t|z|^2z\right)\partial_z\nonumber \\
          &&+\left(-\frac{1}{4}iz+\frac{1}{2}t\bar{z}
             +\frac{1}{96}iz^3-\frac{1}{48}i\bar{z}^3
             +\frac{1}{16}i|z|^2z
             -\frac{1}{8}t|z|^2\bar{z}\right)\partial_{\bar{z}}.
             \label{eq:2-49}
\end{eqnarray}
The corresponding Lie equation or renormailization group
equation reads
\begin{eqnarray}
         \partial_\varepsilon z &
             =&\frac{1}{4}i\bar{z}+\frac{1}{2}tz
              +\frac{1}{48}iz^3-\frac{1}{96}i\bar{z}^3
              -\frac{1}{16}i|z|^2\bar{z}
              -\frac{1}{8}t|z|^2z.
       \label{eq:2-50}
\end{eqnarray}
This renormalization group equation is also consistent with the naive
expanded solution of (\ref{eq:2-46}),\cite{Naiveexpansion}
\begin{eqnarray}
        z&=&Ae^{i(t+\theta)}\nonumber \\
         &&
	+\varepsilon\Biggl\{
         \frac{A}{2}\left(1-\frac{A^2}{4}\right)ite^{-i(t+\theta)}
                 +\frac{A}{4}\left(1-\frac{A^2}{4}\right)ie^{i(t+\theta)}
              \nonumber\\  
         &&\hspace{4cm}       +\frac{A^3}{48}ie^{-3i(t+\theta)}
                 +\frac{A^3}{96}ie^{3i(t+\theta)}
           \Biggr\}.
       \label{eq:2-53}
\end{eqnarray}
Defining  $z_0:=Ae^{-i(t+\theta)}$, Eq.(\ref{eq:2-53})
reads
\begin{eqnarray}
       z&=&z_0+\varepsilon\left\{
           \frac{1}{4}i\bar{z}_0+\frac{1}{2}tz_0
           +\frac{1}{48}iz_0^3+\frac{1}{96}i{\bar{z}}_0^3
           -\frac{1}{16}i|z_0|^2\bar{z}_0-\frac{1}{8}t|z_0|^2z_0
          \right\},
       \label{eq:2-54}
\end{eqnarray}
which is the Euler scheme difference equation corresponding to the
renormalization equation
(\ref{eq:2-50}).
The renormalization group equation  (\ref{eq:2-50}) is approximated
for large $t$ by
\begin{eqnarray}
         \partial_\varepsilon z =
             \frac{1}{2}tz\left(1-\frac{1}{4}|z|^2\right).
       \label{eq:2-55}
\end{eqnarray}
Under the transformation of coordinates $(z,\bar{z}) \mapsto
(A,\alpha)$, with
\begin{eqnarray}
	    A:=|z|,\ \
            \alpha:=\frac{i}{2}{\rm Log}\left(\frac{\bar{z}}{z}\right),
        \label{eq:2-56}
\end{eqnarray}
Eq.(\ref{eq:2-55}) becomes
\begin{eqnarray}
      \left\{
       \begin{array}{lll}
        \partial_\varepsilon A &=&
            \frac{A}{2}\left(1-\frac{A^2}{4}\right)t,\\
        \partial_\varepsilon \alpha &=& 0.
       \end{array}
       \right.
        \label{eq:2-57}
\end{eqnarray}
or
\begin{eqnarray}
      \left\{
       \begin{array}{lll}
        \partial_\tau A &=&
            \frac{A}{2}\left(1-\frac{A^2}{4}\right),\\
        \partial_\tau \alpha &=& 0,
       \end{array}
       \right.
\end{eqnarray}
where $\tau:=\varepsilon t$.
This renormalization group equation is equivalent to that
obtained with the conventional renormalization group
method \cite{IllinoisRG}.

Because the determining equation of the Lie symmetry is linear in each
component of the infinitesimal generator, it is easy to calculate
higher-order corrections to the leading-order Lie symmetry.  The
determining equation to order $\varepsilon^1$ is 
\begin{eqnarray}
&& {\eta^z}^{(1)}|_{\scriptsize
z'+iz-\varepsilon\left(\frac{1}{2}z-\frac{1}{2}\bar{z}
               +\frac{1}{24}z^3-\frac{1}{8}|z|^2z
               +\frac{1}{8}|z|^2\bar{z}-\frac{1}{24}\bar{z}^3\right)}+i\eta^z
            \nonumber \\
&&             -\varepsilon\left(
                   \frac{1}{2}z-\frac{1}{2}\bar{z}
                  +\frac{1}{24}z^3-\frac{1}{8}|z|^2z
                  +\frac{1}{8}|z|^2\bar{z}-\frac{1}{24}\bar{z}^3
                 \right)=O(\varepsilon ^2)   \label{eq:2-58}
\end{eqnarray}
The solution of (\ref{eq:2-58}) gives the following approximate symmetry to 
order $\varepsilon^1$
   for large $t$:
\begin{eqnarray}
     \eta^z = \frac{1}{2}tz\left(1-\frac{1}{4}|z|^2\right)
              +\varepsilon\frac{1}{4}itz
                  \left(1+\frac{1}{4}|z|^2
                        -\frac{1}{16}|z|^4\right).
\end{eqnarray}
Then, we obtain the  renormalization group equation to order $\varepsilon^2$,
\begin{eqnarray}
    \frac{dz}{d\varepsilon}
          = \frac{1}{2}z\left(1-\frac{1}{4}|z|^2\right)t
           +\varepsilon\frac{1}{4}iz
                   \left(1+\frac{1}{4}|z|^2
                         -\frac{1}{16}|z|^4\right)t,
\end{eqnarray}
or
\begin{eqnarray}
    \frac{dz}{d\tau}
          = \frac{1}{2}z\left(1-\frac{1}{4}|z|^2\right)
            +\varepsilon\frac{1}{4}iz\left(1+\frac{1}{4}|z|^2
                             -\frac{1}{16}|z|^4\right).
\end{eqnarray}

\section{Concluding remarks}
        \label{sec4}
   In the method presented here, the renormalization equation appears as
   the main ingredient in the Lie group theory. More specifically, 
   we find that the Lie equation for the Lie point group in the expanded
   space that includes the perturbation parameter is the renormalization
   group equation. In the conventional RG method, it is necessary  to
   calculate naive perturbation solutions of a given nonlinear system,
   which is often tedious, due to the nonlinearity of the system.
   To avoid this step, the proto-RG approach was proposed in the
   perturbation analysis of a perturbed system \cite{NozakiOono}. The
   present method does not require any pertubational analyses to determine
   approximate solutions of a perturbed system but, instead, approximate
   symmetries. Because the determinig equation is always a linear system
   for the symmetry, it is easier to calculate an approximate symmetry
   than to obtain an approximate solution of the nonlinear system. Thus,
   the present method always remains within the framework of Lie group
   theory and completely frees us from the need to calculate naive perturbation
   solutions. 

   It should be noted that the present method can be applied not only to
   continuous  systems but also to discrete systems, owing to the general
   framework of the Lie group.


%

\end{document}